\documentclass[twocolumn,english,superscriptaddress,citeautoscript,showpacs,preprintnumbers,amsmath,amssymb,prb,floatfix,footinbib]{revtex4}
\usepackage[scaled=2]{helvet}
\usepackage[T1]{fontenc}
\usepackage[utf8]{luainputenc}
\usepackage{textcomp}
\usepackage{amsmath}
\usepackage{amssymb}
\usepackage{graphicx}
\usepackage{subscript}

\makeatletter

\providecommand{\tabularnewline}{\\}

\@ifundefined{textcolor}{}
{%
 \definecolor{BLACK}{gray}{0}
 \definecolor{WHITE}{gray}{1}
 \definecolor{RED}{rgb}{1,0,0}
 \definecolor{GREEN}{rgb}{0,1,0}
 \definecolor{BLUE}{rgb}{0,0,1}
 \definecolor{CYAN}{cmyk}{1,0,0,0}
 \definecolor{MAGENTA}{cmyk}{0,1,0,0}
 \definecolor{YELLOW}{cmyk}{0,0,1,0}
}

\usepackage[scaled=2]{helvet}\usepackage{amstext}

\makeatletter%%%%%%%%%%%%%%%%%%%%%%%%%%%%%% Textclass specific LaTeX commands.
\@ifundefined{textcolor}{}{\definecolor{BLACK}{gray}{0}\definecolor{WHITE}{gray}{1}\definecolor{RED}{rgb}{1,0,0}\definecolor{GREEN}{rgb}{0,1,0}\definecolor{BLUE}{rgb}{0,0,1}\definecolor{CYAN}{cmyk}{1,0,0,0}\definecolor{MAGENTA}{cmyk}{0,1,0,0}\definecolor{YELLOW}{cmyk}{0,0,1,0}}

\usepackage{dcolumn}% Align table columns on decimal point
\usepackage{bm}% bold math
\usepackage{times}% font that prx use
\usepackage{hyperref}
\hypersetup{colorlinks=true,linkcolor=red,citecolor=blue}
\makeatother

\usepackage{babel}

\makeatother

\usepackage{babel}
\begin{document}

\title{Magnetic ground state of superconducting Eu(Fe$_{0.88}$Ir$_{0.12}$)$_{2}$As$_{2}$:
a combined neutron diffraction and first-principles calculation study}

\author{W. T. Jin}

\email{w.jin@fz-juelich.de}

\affiliation{Jülich Centre for Neutron Science JCNS and Peter Grünberg Institut
PGI, JARA-FIT, Forschungszentrum Jülich GmbH, D-52425 Jülich, Germany}

\affiliation{Jülich Centre for Neutron Science JCNS, Forschungszentrum Jülich
GmbH, Outstation at MLZ, Lichtenbergstraße 1, D-85747 Garching, Germany}

\author{Wei Li}

\affiliation{State Key Laboratory of Functional Materials for Informatics and
Shanghai Center for Superconductivity, Shanghai Institute of Microsystem
and Information Technology, Chinese Academy of Sciences, Shanghai
200050, China }

\author{Y. Su}

\affiliation{Jülich Centre for Neutron Science JCNS, Forschungszentrum Jülich
GmbH, Outstation at MLZ, Lichtenbergstraße 1, D-85747 Garching, Germany}

\author{S. Nandi}

\affiliation{Jülich Centre for Neutron Science JCNS and Peter Grünberg Institut
PGI, JARA-FIT, Forschungszentrum Jülich GmbH, D-52425 Jülich, Germany}

\affiliation{Jülich Centre for Neutron Science JCNS, Forschungszentrum Jülich
GmbH, Outstation at MLZ, Lichtenbergstraße 1, D-85747 Garching, Germany}

\author{Y. Xiao}

\affiliation{Jülich Centre for Neutron Science JCNS and Peter Grünberg Institut
PGI, JARA-FIT, Forschungszentrum Jülich GmbH, D-52425 Jülich, Germany}

\author{W. H. Jiao}

\affiliation{School of Science, Zhejiang University of Science and Technology, Hangzhou 310023, China}

\author{M. Meven}

\affiliation{RWTH Aachen University, Institut für Kristallographie, D-52056 Aachen,
Germany}

\affiliation{Jülich Centre for Neutron Science JCNS, Forschungszentrum Jülich
GmbH, Outstation at MLZ, Lichtenbergstraße 1, D-85747 Garching, Germany}

\author{A. P. Sazonov}

\affiliation{RWTH Aachen University, Institut für Kristallographie, D-52056 Aachen,
Germany}

\affiliation{Jülich Centre for Neutron Science JCNS, Forschungszentrum Jülich
GmbH, Outstation at MLZ, Lichtenbergstraße 1, D-85747 Garching, Germany}

\author{E. Feng}

\affiliation{Jülich Centre for Neutron Science JCNS, Forschungszentrum Jülich
GmbH, Outstation at MLZ, Lichtenbergstraße 1, D-85747 Garching, Germany}

\author{Yan Chen}

\affiliation{Department of Physics, State Key Laboratory of Surface Physics and
Laboratory of Advanced Materials, Fudan University, Shanghai 200433,
China }

\author{C. S. Ting}

\affiliation{Texas Center for Superconductivity and Department of Physics, University
of Houston, Houston, Texas 77204, USA}

\author{G. H. Cao}

\affiliation{Department of Physics, Zhejiang University, Hangzhou 310027, China}

\author{Th. Brückel}

\affiliation{Jülich Centre for Neutron Science JCNS and Peter Grünberg Institut
PGI, JARA-FIT, Forschungszentrum Jülich GmbH, D-52425 Jülich, Germany}

\affiliation{Jülich Centre for Neutron Science JCNS, Forschungszentrum Jülich
GmbH, Outstation at MLZ, Lichtenbergstraße 1, D-85747 Garching, Germany}

\date{\today}% It is always \today, today,

\begin{abstract}
The magnetic order of the localized Eu$^{2+}$ spins in optimally-doped Eu(Fe$_{1-x}$Ir$_{x}$)$_{2}$As$_{2}$ ($\mathit{x}$ = 0.12) with superconducting transition temperature $\mathit{T_{SC}}$ = 22 K was investigated by single-crystal neutron diffraction. The Eu$^{2+}$ moments were found to be ferromagnetically aligned along the $\mathit{c}$-direction with an ordered moment of 7.0(1) $\mu_{B}$ well below the magnetic phase transition temperature $\mathit{T_{C}}$ = 17 K. No evidence of the tetragonal-to-orthorhombic structural phase transition was found in this compound within the experimental uncertainty, in which the spin-density-wave (SDW) order of the Fe sublattice is supposed to be completely suppressed and the superconductivity gets fully developed. The ferromagnetic groud state of the Eu$^{2+}$ spins in Eu(Fe$_{0.88}$Ir$_{0.12}$)$_{2}$As$_{2}$ was supported by the first-principles density functional calculation. In addition, comparison of the electronic structure calculations between Eu(Fe$_{0.875}$Ir$_{0.125}$)$_{2}$As$_{2}$ and the parent compound EuFe$_{2}$As$_{2}$ indicates stronger hybridization and more expanded bandwith due to the Ir substitution, which together with the introduction of electrons might work against the Fe-SDW in favor of the superconductivity. 

\end{abstract}

\pacs{71.15.Mb, 74.70.Xa, 75.25.-j, }% PACS, the Physics and Astronomy

\maketitle

\section{Introduction}

The discovery of unconventional superconductivity in the iron pnictides in 2008 \cite{Kamihara_08} has provided a new opportunity to study the intriguing interplay between superconductivity and magnetism, as the superconductivity in the Fe-based superconductors was again found to emerge in close proximity to the magnetic instability,\cite{Cruz_08,Dai_12} similar to that in cuprates and heavy-fermion compounds. Among various classes of Fe-based superconductors, the ternary ``122'' family, $\mathit{A}$Fe$_{2}$As$_{2}$ ($\mathit{A}$ = Ba, Sr, Ca, etc) has attracted more attention due to the relative high superconducting transition temperature ($\mathit{T_{SC}}$) and the ease in obtaining large, high-quality single crystals.

EuFe$_{2}$As$_{2}$ is a unique member of the ``122'' family as it contains two magnetic sublattices. The $\mathit{A}$ site is occupied by an $\mathit{S}$-state rare-earth Eu$^{2+}$ ion possessing a 4$\mathit{f}$$^{7}$ electronic configuration with an electron spin $\mathit{S}$ = 7/2, corresponding to a theoretical effective magnetic moment of 7.94 $\mathit{\mu_{B}}$.\cite{Marchand_78} This compound undergoes a spin-density-wave (SDW) transition in the Fe sublattice concomitant with a tetragonal-to-orthorhombic structural
phase transition below 190 K. In addition, the localized Eu$^{2+}$ spins order in an A-type antiferromagnetic (A-AFM) structure (ferromagnetic layers ordering antiferromagnetically along the $\mathit{c}$ direction) below 19 K.\cite{Herrero-Martin_09,Xiao_09,Jiang_09_NJP} As in other iron pnictides, superconductivity can be achieved in the EuFe$_{2}$As$_{2}$ family when the structural distortion and the SDW order of Fe are significantly suppressed by either chemical substitution  \cite{Jeevan_08,Ren_09,Jiang_09,Jiao_11,Jiao_13} or application of external pressure.\cite{Miclea_09,Terashima_09} However, there is no clear picture so far regarding how the magnetic order of the Eu$^{2+}$spins evolves with doping or pressure and how it is linked to the superconductivity. Recently, by means of neutron diffraction and resonant magnetic x-ray scattering, the authors have determined the magnetic structure of superconducting Eu(Fe$_{1-x}$Co$_{x}$)$_{2}$As$_{2}$ ($\mathit{x}$ = 0.18, $\mathit{T_{SC}}$ = 8 K ) \cite{Jin_13} and EuFe$_{2}$(As$_{1-x}$P$_{x}$)$_{2}$ ($\mathit{x}$ = 0.15 and 0.19, $\mathit{T_{SC}}$ = 25 K and 27 K, respectively).\cite{Nandi_14,Nandi_14_neutron} In both systems, the Eu$^{2+}$spins order ferromagnetically in the superconducting state. The difference between the two systems mentioned above is that in P-doped EuFe$_{2}$As$_{2}$, the superconducting transition temperature$\mathit{T_{SC}}$ is higher than the Curie temperature $\mathit{T_{C}}$ (19 K), while in the Co-doped compound, the sequence is inversed with $\mathit{T_{C}}$ (17 K) higher than $\mathit{T_{SC}}$. The coexistence of ferromagnetism and superconductivity, two antagonistic collective phenomena, makes the doped EuFe$_{2}$As$_{2}$ system quite striking and more attentions are being attracted onto the exploration for exotic superconductivity within this family. 

Recently, superconductivity was observed in $\mathit{\textrm{5}d}$ transition metal element doped Eu(Fe$_{1-x}$Ir$_{x}$)$_{2}$As$_{2}$ with $\mathit{T_{SC}}$ up to \textasciitilde{} 22 K.\cite{Paramanik_13,Jiao_13} However, the magnetic structure of the  Eu$^{2+}$ moments near the optimal Ir-doping level remains quite controversial.\cite{Paramanik_13,Jiao_13} Based on macroscopic measurements on pollycrystalline samples, Paramanik $et\, al.$ \cite{Paramanik_13} proposed a canted antiferromagnetic structure with a ferromagnetic component as the magnetic ground state of the Eu$^{2+}$ spins for Eu(Fe$_{0.86}$Ir$_{0.14}$)$_{2}$As$_{2}$ with $\mathit{T_{SC}}$ = 22.5 K. Meanwhile, Jiao $et\, al.$ \cite{Jiao_13} concluded a ferromagnetic groud state for  Eu(Fe$_{0.88}$Ir$_{0.12}$)$_{2}$As$_{2}$ with $\mathit{T_{SC}}$ = 22 K based on similar measurements using a single crystal and further proposed a temperature-induced spin-reorientation scenario, in which the Eu$^{2+}$ spins tilt from $\mathit{ab}$ plane
to the $\mathit{c}$-axis while cooling. Thus, it is important to unambiguously determine the real magnetic ground state of the Eu$^{2+}$ spins in optimally-doped Eu(Fe$_{1-x}$Ir$_{x}$)$_{2}$As$_{2}$ by neutron diffraction, the preferred experimental method for the
bulk probe of the magnetic order. However, due to the large neutron absorption cross sections of both Eu and Ir, the neutron diffraction measurement on such material is quite challenging. Nevertheless, by significant reduction of the absorption effect using hot neutrons,
such measurements were proved to be feasible for platelike crystals of good quality according to our previous experiences on similar Eu-containing iron pnictides.\cite{Xiao_09,Xiao_10,Jin_13} Here we present the results of our neutron diffraction measurements on a high-quality
Eu(Fe$_{1-x}$Ir$_{x}$)$_{2}$As$_{2}$ ($\mathit{x}$ = 0.12) single crystal, which is at the optimal Ir-doping level for superconductivity ($\mathit{T_{SC}}$ = 22 K). The magnetic ground state of the Eu$^{2+}$ spins is revealed to be a ferromagnetic alignment along the $\mathit{c}$-direction. This experimental result is supported by first-principles magnetic structure calculations. We do not find any evidence for the temperature-induced spin-canting scenario of the Eu$^{2+}$ moments. In addition, no evidence suggesting the existence of a structural phase transition is observed. By comparison with the Co-doped EuFe$_{2}$As$_{2}$, the role of $5\mathit{d}$ Ir is revealed to be more effective in introducing robust superconductivity, which might be correlated to the broadening of the bands and increasing hybridization caused by the Ir substitution, as suggested by the band structure calculations.

\section{Experimental Details and Theoretical Methods}

Single crystals of Eu(Fe$_{1-x}$Ir$_{x}$)$_{2}$As$_{2}$ ($\mathit{x}$ = 0.12) were grown from self-flux (Fe, Ir)As.\cite{Jiao_13} The as-grown crystals could be easily cleaved. The $\mathit{c}$ axis is perpendicular to their surfaces, as confirmed by x-ray diffraction. The chemical composition of the crystals was determined by energy dispersive x-ray (EDX) analysis. A 24 mg platelike single crystal with dimensions \textasciitilde{} 4 $\times$ 4 $\times$ 0.5 mm$^{3}$ was selected for the neutron diffraction measurements, which were performed on the hot-neutron four-circle diffractometer HEIDI at Heinz Maier-Leibnitz Zentrum (MLZ), Garching (Germany).\cite{Matin_07} A Ge (4 2 2) monochromator was chosen to produce a monochromatic neutron beam with the wavelength of 0.793 \AA, for which the neutron absorption cross section of Eu and Ir is reduced to 1998 and 187 barns, respectively. An Er filter was used to minimize the $\lambda/2$ contamination. The single-crystal sample was mounted on a thin aluminum holder with a small amount of GE varnish and put inside a standard closed-cycle cyrostat. The diffracted neutron beam was collected with a $^{3}$He single detector. The integrated intensities of 508 (154 independent) reflections at 25 K (above the magnetic ordering temperature of the Eu$^{2+}$ moments) and 478 (145 independent) reflections at 2.5 K, respectively, were collected via rocking-curve scans. The obtained reflection sets at both temperatures were normalized to the monitor and corrected by the Lorentz factor. The DATAP program was used for the absorption correction by considering the size and shape of the crystal.\cite{Coppens_65} Refinement of both nuclear and magnetic structures was carried out using the FULLPROF program suite.\cite{Rodriguez_93} For macroscopic characterizations, a platelike crystal of 12.9 mg from the same batch was used. The resistivity
and magnetization were measured using a Quantum Design physical property measurement system (PPMS) and a Quantum Design magnetic property measurement system (MPMS), respectively. 

The first-principles calculations presented in this work were performed using the projected augmented-wave method,\cite{Blochl_94} as implemented in the VASP code.\cite{Kresse_96} The exchange-correlation potential was calculated using the generalized gradient approximation (GGA) as proposed by Pedrew, Burke, and Ernzerhof.\cite{Perdew_96} We have included the strong Coulomb repulsion in the Eu-4$\mathit{f}$ orbitals on a mean-field level using the GGA+$\mathit{U}$ approximation. There exist no spectroscopy data for EuFe$_{2}$As$_{2}$ and Eu(Fe$_{1-x}$Ir$_{x}$)$_{2}$As$_{2}$. Therefore, throughout this work, we have used a $\mathit{U}$ of 8 eV,\cite{LiW_12} which is the standard value for an Eu$^{2+}$ ion. The results were checked for consistency with varying $\mathit{U}$
values. We did not apply $\mathit{U}$ to the itinerant Fe $3\mathit{d}$ orbitals. Additionally, the spin-orbit coupling is included for all atoms with the second variational method in the calculations. These calculations were performed using the experimental crystal structure,
as determined by the neutron diffraction measurements, while all the atomic positions were optimized until the largest force on each atom was 0.005 eV/\AA).

\section{Results }

\subsection{Macroscopic characterizations}

The temperature dependence of the normalized in-plane resistivity ($\rho_{ab})$ of the Eu(Fe$_{0.88}$Ir$_{0.12}$)$_{2}$As$_{2}$ single crystal is shown in Fig. 1. The resisitivity decreases linearly while cooling and no anomaly associated with possible phase transitions
is observed until a sharp superconducting transition occurs at $\mathit{T_{SC}}$ = 22 K. The zero-resistance state is achieved below 20.5 K, as illustrated in the inset of Fig. 1. The superconducting transition temperature of this sample is the achievable maximal value in the series of Eu(Fe$_{1-x}$Ir$_{x}$)$_{2}$As$_{2}$ single crystals. Therefore we refer this sample with $\mathit{x}$ = 0.12 as the optimally-doped one. The optimal $\mathit{T_{SC}}$ around 22 K is quite close to that of the polycrystalline Eu(Fe$_{1-x}$Ir$_{x}$)$_{2}$As$_{2}$ (22.6 K for $\mathit{x}$ = 0.14) \cite{Paramanik_13} but the reentrant behavior of the resistivity reported there is not observed in our single crystal sample. 

\begin{figure}
\centering{}\includegraphics{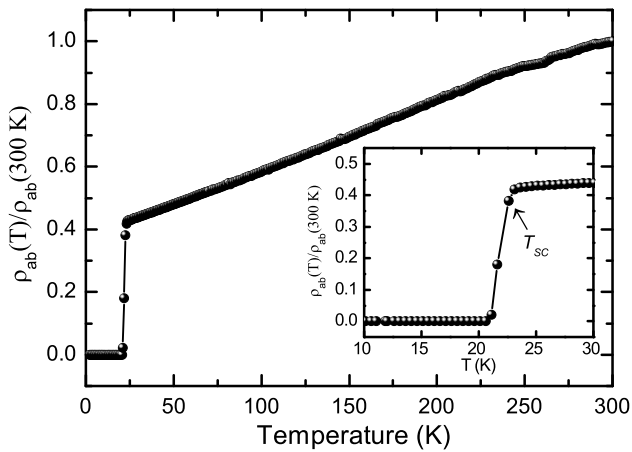}

\caption{The temperature dependence of the normalized in-plane resistivity ($\rho_{ab})$ of the Eu(Fe\textsubscript{0.88}Ir\textsubscript{0.12})\textsubscript{2}As\textsubscript{2} single crystal. The inset gives an enlarged view of the \textit{$\rho$-T} curve around $T_{SC}$.}
\end{figure}

Figure 2 shows the temperature dependence of the volume magnetic susceptibility ($\chi_{v}$) of Eu(Fe$_{0.88}$Ir$_{0.12}$)$_{2}$As$_{2}$ under an applied field of 10 Oe along the $\mathit{c}$-direction of the crystal. A distinct diamagnetic response associated with the superconducting transition appears below $\mathit{T_{SC}}$ = 22 K for the zero-field-cooling (ZFC) susceptibility, consistent with the sudden drop in the \textit{$\rho$-T} curve. With further cooling the diamagnetic signal is weakened by the onset of the ferromagnetic order of the Eu$^{2+}$ spins around 17 K ($\mathit{T_{C}}$) as revealed by our neutron measurements presented below, where the ZFC susceptibility reaches a local maximum. When the temperature is further decreased, the superconductivity wins over the ferromagnetism of the Eu sublattice and the ZFC susceptibility decreases again. The absence of the Meissner state as shown in the field-cooling (FC) susceptibility seems a common feature in various superconducting Eu-based ``122'' compounds,\cite{Jeevan_11,Jiao_11,Jiao_13} probably due to the very strong internal field produced by the ferromagnetism of the localized Eu$^{2+}$ moments. Detailed macroscopic measurements on the same crystal have confirmed the bulk property of the superconductivity.\cite{Jiao_13}

\begin{figure}
\centering{}\includegraphics{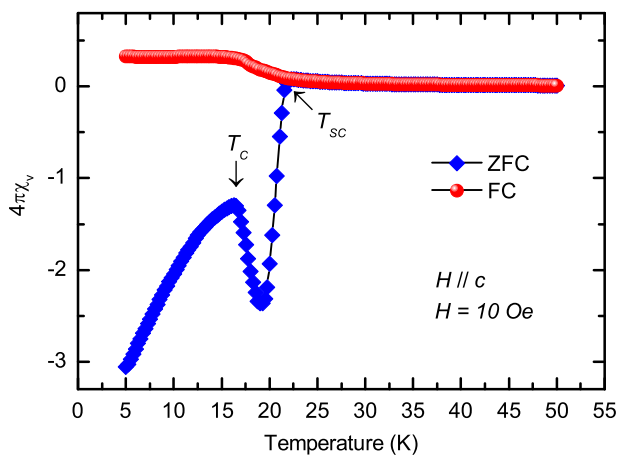}

\caption{The temperature dependence of the volume magnetic susceptibility ($\chi_{v}$) of Eu(Fe\textsubscript{0.88}Ir\textsubscript{0.12})\textsubscript{2}As\textsubscript{2 }measured in an applied field of 10 Oe along the \textit{c}-direction in a ZFC and FC process, respectively. }

\end{figure}

\subsection{Neutron diffraction }

Previous single-crystal neutron diffraction measurements have revealed that the parent compound EuFe$_{2}$As$_{2}$ undergoes a structural phase transition (SPT) from tetragonal (space group $\mathit{I4/mmm}$) to orthorhombic ($\mathit{Fmmm}$) below $\mathit{T_{S}}$ = 190 K.\cite{Xiao_09} With chemical doping, the transition temperature $\mathit{T_{S}}$ is suppressed in favor of the occurrence of superconductivity.\cite{Jin_13,Nandi_14} In order to clarify the presence or absence of such SPT in the superconducting
Eu(Fe$_{0.88}$Ir$_{0.12}$)$_{2}$As$_{2}$, the rocking-curve scan of the (2 2 0)$_{T}$ reflection (in the tetragonal notation), which is most sensitive to the in-plane structural distortion, was performed at different temperatures while cooling. Figure 3(a) shows the temperature dependencies of both the integrated intensity and the full-width-at-half-maximum (FWHM) of the (2 2 0)$_{T}$ peak together with those of the (0 0 8)$_{T}$ peak, which is shown for comparison. It is evident that the integrated intensity of the (2 2 0)$_{T}$ peak evolves smoothly, without showing any anomaly related to the orthorhombic distortion.\cite{Lester_09,Jin_13} Although the (2 2 0)$_{T}$ peak broadens while cooling, it behaves in a very similar way as the (0 0 8)$_{T}$ peak, indicating that all the reflections uniformly broaden while cooling and no tetragonal-to-orthorhombic SPT can be identified. This might be due to slight bending of the crystal during the cooling process. To be more confident about the absence of the SPT, two dimensional Q scans in the orthorhombic (H K 0)$\mathit{_{O}}$ plane were performed at 5.5 K and shown in Fig. 3(b). Only a single peak centered at (4 0 0)$_{O}$ or (2 2 0)$_{T}$ can be observed. The diffuse ring appearing on the high-Q side of the (4 0 0)$_{O}$ peak is due to the reflection from the aluminum
sample holder. The absence of the SPT in superconducting Eu(Fe$_{0.88}$Ir$_{0.12}$)$_{2}$As$_{2}$ is well consistent with the linear temperature dependence of the in-plane resistivity above $\mathit{T_{SC}}$ as shown in Fig. 1, from which no change in the Fermi surface nesting is expected. This is in stark contrast to the case of the 3$\mathit{d}$ Co-doped superconducting Eu(Fe$_{0.82}$Co$_{0.18}$)$_{2}$As$_{2}$, where the SPT was revealed by neutron diffraction to occur around 90 K, the temperature where a pronounced kink in the the \textit{$\rho_{ab}$-T} curve can be observed.\cite{Jin_13} Since the SDW order of the Fe$^{2+}$ moments generally follows the occurence of the SPT closely, it is unlikely that the antiferromagnetism of Fe develops in Eu(Fe$_{0.88}$Ir$_{0.12}$)$_{2}$As$_{2}$.
Here both the SPT and the Fe-SDW order are believed to be completely suppressed.

\begin{figure}
\centering{}\includegraphics{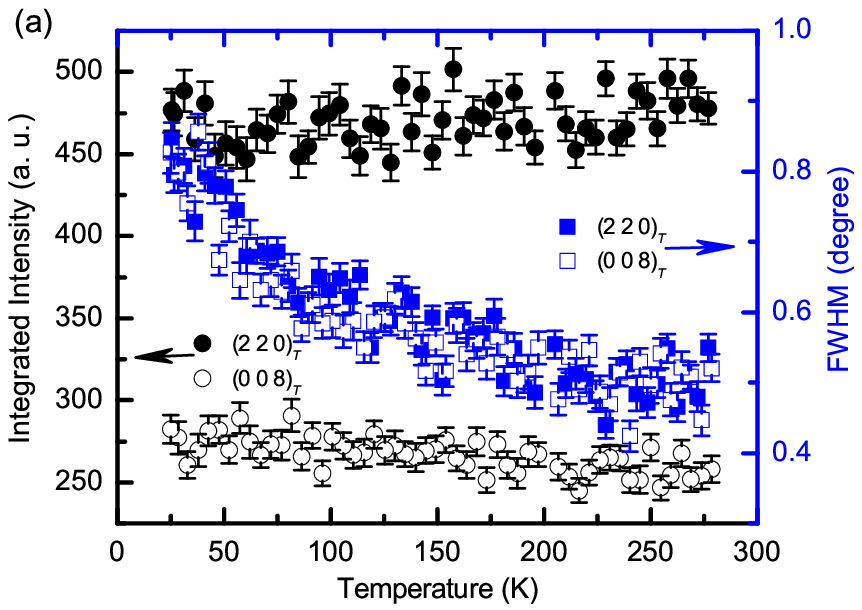}
\includegraphics{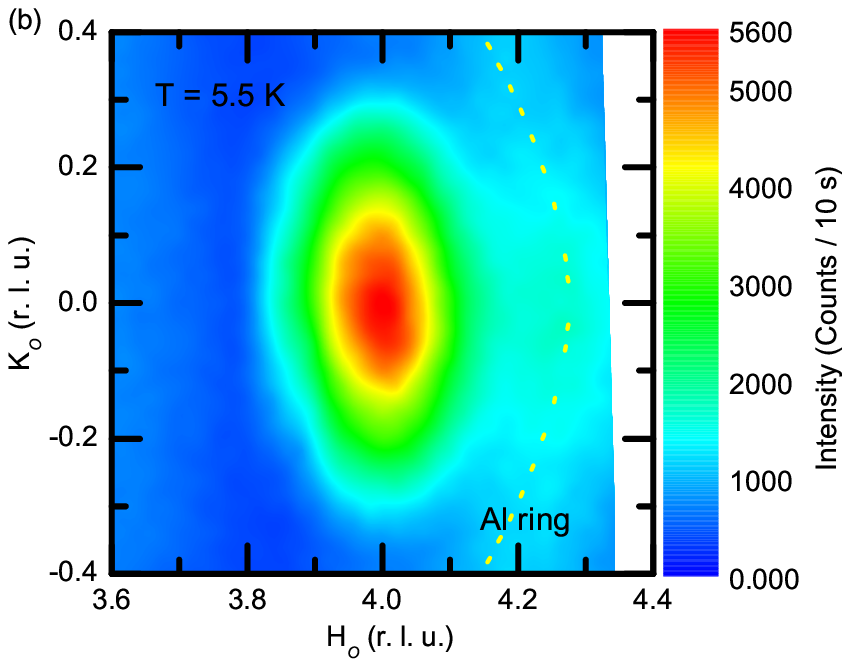}

\caption{(a) The temperature dependencies of the integrated intensity (black circles) and the peak width (FWHM, blue squares) of the (2 2 0)$_{T}$ and (0 0 8)$_{T}$ reflections. (b) The contour map of (4 0 0)$_{O}$ or (2 2 0)$_{T}$ reflection at $\mathit{T}$ = 5.5 K, which confirms the absence of the SPT transition while cooling within our experimental resolution. The conversion of Miller indices between the orthorhombic and tetragonal notations is $\mathit{H_{T}}=(H_{O}$ + $K_{O})/2$, $\mathit{K_{T}}=(H_{O}$ - $K_{O})/2$ and $L_{T}$ = $L_{O}.$ The diffuse ring appearing on the high-Q side of the (4 0 0)$_{O}$ peak is a powder ring from the aluminum sample holder.}
\end{figure}

To conclude about the magnetic ground state of the Eu$^{2+}$ moments, rocking curve scans of several representative reflections were performed at both 25 K and at the base temperature, which are above and well below the magnetic transition temperature of the Eu sublattice, respectively. As shown in Fig. 4(a) and (b), the weak nuclear reflections at 25 K, (1 1 0)$_{T}$ and (1 0 1)$_{T}$, are remarkably enhanced at the base temperature, indicating a huge ferromagnetic contribution from the Eu$^{2+}$ spins. On the other hand, the (0 0 6)$_{T}$ reflection shows no discernible change upon cooling {[}Fig. 4(c){]}, suggesting that the ferromagnetic component of the Eu$^{2+}$ spins in the $\mathit{ab}$ plane is almost zero or can't be resolved within the experimental uncertainty. \cite{Footnote} In other words, within our experimental uncertainty, the Eu$^{2+}$ spins are ferromagnetically aligned along the $\mathit{c}$ direction in the ground state.\cite{Jiao_13} The temperature dependence of the integrated intensity of the (1 1 0)$_{T}$ reflection is plotted in Fig. 4(d). Fitting of the order parameter using the power law $\mathit{I}$ - $\mathit{I_{0}}$ $\propto$($\mathit{T}$ - $\mathit{T_{C}}$)$^{2\beta}$ close to the transition yields the ferromagnetic transition temperature $\mathit{T_{C}}=16.89(7)$ K and the exponent $\mathit{\beta}$ = 0.31(2), close to the critical exponent of the three-dimensional Ising model ($\mathit{\beta}$ = 0.326). $\mathit{T_{C}}$ determined here is in good agreement with the value from the magnetization measurement (Fig. 2). In addition, in Ref. \onlinecite{Jiao_13}, a temperature-induced spin-reorientation scenario was proposed for the same compound, in which the ferromagnetic Eu$^{2+}$ spins flop from the $\mathit{c}$-direction into the $\mathit{ab}$ plane when the temperature is between 17.4 K and 20 K. According to our observation, such preceding in-plane ferromagnetism, if indeed developed, can not be long-range ordered, since the integrated intensity of the (0 0 6)$_{T}$ reflection, which is most sensitive to the in-plane, long-range ordered ferromagnetic component, remains almost constant below 25 K. The magnetic moment of the Eu$^{2+}$ spins is pinned along the $\mathit{c}$-axis below $\mathit{T_{C}}$ based on our measurements. However, the possibility of a short-range or fluctuating in-plane ferromagnetism above $\mathit{T_{C}}$ can't be ruled out.

\begin{figure}
\centering{}\includegraphics{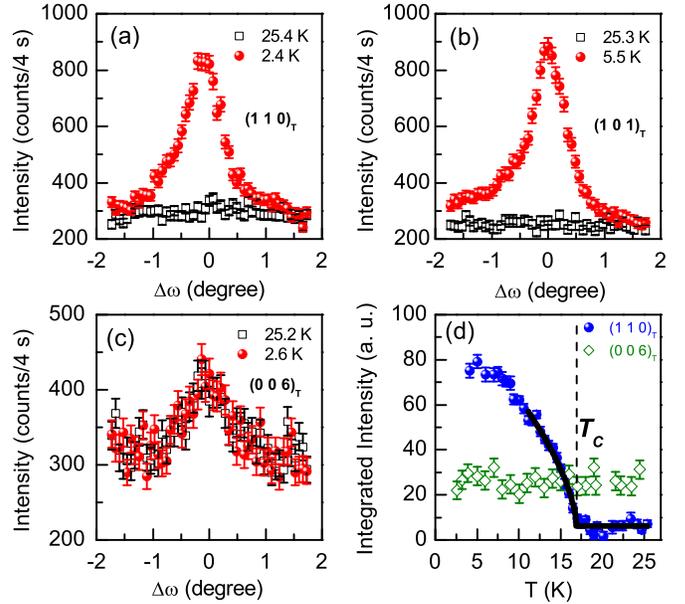}

\caption{Rocking curve scans ($\mathit{\omega}$-scans) of (a) (1 1 0)$_{T}$, (b) (1 0 1)$_{T}$ and (c) (0 0 6)$_{T}$ reflections at 25 K and at base temperature, respectively. (d) The temperature dependencies of the integrated intensities of the (1 1 0)$_{T}$ and (0 0 6)$_{T}$. The solid line represents a fit of the ferromagnetic order parameter close to the transition using a power law. The vertical dashed line denotes the ferromagnetic transition temperature, $\mathit{T_{C}}$.}
\end{figure}

Furthermore, two Q scans along the (1 0 L)$_{T}$ and (1 1 L)$_{T}$ directions were performed at base temperature, as shown in Fig. 5(a) and Fig. 5(b), respectively. No magnetic peaks corresponding to the antiferromagnetic order of the Eu$^{2+}$ moments occuring in the
parent compound are observed at (1 0 0)$_{T}$ and (1 1 1)$_{T}$, excluding the existence of any significant amount of undoped or underdoped impurity in the crystal. The magnetic contribution superimposed on the nuclear peak positions, again suggests a ferromagnetic ground
state for the Eu$^{2+}$ spins with the magnetic propagation vector $\mathbf{\mathbf{\mathit{\mathbf{k}}}}=$(0 0 0).

\begin{figure}
\centering{}\includegraphics{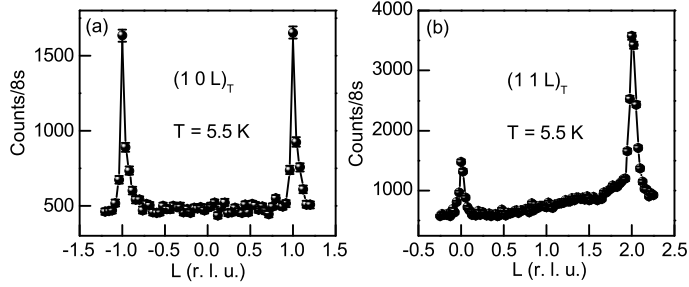}

\caption{Q scans along the L direction through the (1 0 L)$_{T}$ and (1 1 L)$_{T}$ reflections, respectively, at base temperature.}
\end{figure}

To determine precisely the nuclear and magnetic structures of Eu(Fe$_{0.88}$Ir$_{0.12}$)$_{2}$As$_{2}$, the integrated intensities of 508 reflections at 25 K and 478 reflections at 2.5 K were collected. After the absorption correction procedure, the structures were refined using the FULLPROF program within the $\mathit{I4/mmm}$ space group, since no evidence for the tetragonal-orthorhombic structural phase transition was found in both neutron nor resistivity measurements. The results of the refinements are listed in Table 1. The nuclear structure of Eu(Fe$_{0.88}$Ir$_{0.12}$)$_{2}$As$_{2}$ shows no evident difference between 2.5 and 25 K, and the reflections at 2.5 K could be well refined with addition of a ferromagnetic Eu$^{2+}$ moment of 7.0(1) $\mathit{\mu_{B}}$ purely along the $\mathit{c}$
direction. The calculated intensities of nonequivalent reflections according to the refined nuclear and magnetic model are plotted against those observed in Fig. 6. Considering the difficulty associated with the absorption correction on the irregular-shaped crystal, the calculated
and observed intensities are in good agreement.

\begin{table}
\caption{Refinement results for the nuclear and magnetic structures of Eu(Fe\textsubscript{0.88}Ir\textsubscript{0.12})\textsubscript{2}As\textsubscript{2 }at 2.5 K, and the nuclear structure at 25 K. The atomic positions are as follows: Eu, $4a$ (0, 0, 0); Fe/Ir, $8f$ (0.5, 0, 0.25); As, $8i$ (0, 0, $z$). The occupancies of Fe and Ir atoms were fixed to 88\% and 12\%, respectively, according to the chemical composition determined from EDX. Only the isotropic temperature factors ($\mathit{B}$) of all atoms were refined. (Space group: $\mathit{I4/mmm}$ )}

\begin{ruledtabular} %
\begin{tabular}{cccc}
\multicolumn{2}{c}{Temperature} & 2.5 K & 25 K\tabularnewline
\hline 
$a\,(\textrm{\AA)}$ &  & 3.931(2) & 3.932(2)\tabularnewline
$c\,(\textrm{\AA)}$ &  & 11.89(1) & 11.90(1)\tabularnewline
\hline 
Eu  & $B\,$(\AA)\textsuperscript{2}) & 0.78(1)  & 0.79(5) \tabularnewline
 & magnetic propagation vector $\mathbf{k}$ & (0 0 0) & - \tabularnewline
  & $M_{c}$($\mu_{B})$ & 7.0(1)  & - \tabularnewline
Fe/Ir & $B\,$(\AA)\textsuperscript{2}) & 0.49(2) & 0.56(3) \tabularnewline
As  & $z$ & 0.3619(3)  & 0.3620(2) \tabularnewline
 & $B\,$(\AA)\textsuperscript{2}) & 0.57(3)  & 0.64(3) \tabularnewline
\hline 
$R{}_{F^{2}}$  &  & 8.54 & 7.99 \tabularnewline
$R{}_{wF^{2}}$  &  & 9.52 & 9.48\tabularnewline
$R_{F}$  &  & 7.41 & 7.89\tabularnewline
$\chi^{2}$ &  & 0.99 & 0.85\tabularnewline
\multicolumn{4}{l}{Definitions of the agreement factors: \cite{Fullprof_Manual}}\tabularnewline
\multicolumn{4}{c}{$R_{F^{2}}$ = 100$\frac{\sum_{n}[\mid G_{obs,n}^{2}-\sum_{k}G_{calc,k}^{2}\mid]}{\sum_{n}G_{obs,n}^{2}}$,}\tabularnewline
\multicolumn{4}{c}{$R_{wF^{2}}$ = 100$\sqrt{\frac{\sum_{n}w_{n}(G_{obs,n}^{2}-\sum_{k}G_{calc,k}^{2})^{2}}{\sum_{n}w_{n}G_{obs,n}^{2}}}$,}\tabularnewline
\multicolumn{4}{c}{$R_{F}$ = 100$\frac{\sum_{n}[\mid G_{obs,n}-\sqrt{\sum_{k}G_{calc,k}^{2}}\mid]}{\sum_{n}G_{obs,n}}$,}\tabularnewline
\multicolumn{4}{l}{where the index $\mathit{n}$ runs over the observations and the index
$\mathit{k}$ runs }\tabularnewline
\multicolumn{4}{l}{over the reflections contributing to the observation $\mathit{n}$.}\tabularnewline
\multicolumn{4}{l}{$\mathit{G^{2}}$ is the square of the structure factor. }\tabularnewline
\multicolumn{4}{l}{$\mathit{w_{n}}$= $\mathit{1/\sigma_{n}^{2}}$ is the weight where
$\mathit{\sigma_{n}^{2}}$ is the variance of $\mathit{G_{obs,n}}$.}\tabularnewline
\end{tabular}\end{ruledtabular} 
\end{table}

\begin{figure}
\centering{}\includegraphics{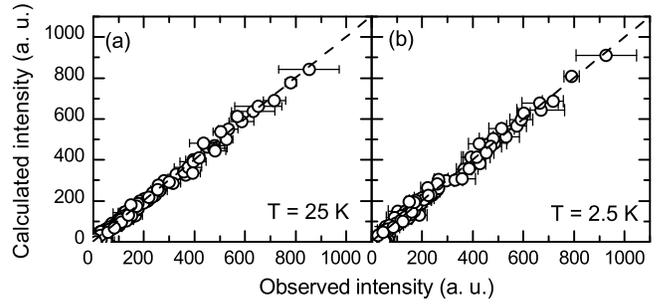}

\caption{The observed and calculated integrated intensities of the nonequivalent reflections at 25 K (a) and 2.5 K (b).}
\end{figure}

The magnetic ground state of the Eu$^{2+}$ spins in superconducting Eu(Fe$_{1-x}$Ir$_{x}$)\textsubscript{2}As\textsubscript{2 }($\mathit{x}$ = 0.12) is quite similar to that in superconducting Eu(Fe$_{1-x}$Co$_{x}$)$_{2}$As$_{2}$ ($\mathit{x}$ = 0.18), \cite{Jin_13} in which the Eu$^{2+}$ moments also order ferromagnetically along the $\mathit{c}$-direction as determined by neutron diffraction. However, both the SPT and the SDW order of Fe occur in the Co-doped compound, while they were completely suppressed in the Ir-doped crystal with the doping level even 6\% lower. Considering the very similar effect in introducing the electron carriers of Co ($\mathit{\textrm{3}d}$$\mathit{^{7}}$$\mathit{\textrm{4}s}$$\mathit{^{2}}$) and Ir ($\mathit{\textrm{5}d}$$^{\mathit{7}}$$\mathit{\textrm{6}s}$$\mathit{^{2}}$), the more effective suppression of the SPT and the Fe-SDW in the Ir-doped EuFe$_{2}$As\textsubscript{2 }can be attributed to the role of more extended $\mathit{\textrm{5}d}$ orbitals for Ir. This will be further discussed below.

\subsection{First-principles calculations}

To better understand the role of $\mathit{\textrm{5}d}$ Ir doping, the electronic structure calculation for Eu(Fe$_{0.875}$Ir$_{0.125}$)$_{2}$As$_{2}$ was performed using supercell method in the quenched paramagnetic state on the Fe layers, in which no spin polarization is allowed on the Fe or Ir ions in the calculations. The density of states (DOS) of Eu(Fe$_{0.875}$Ir$_{0.125}$)$_{2}$As$_{2}$ is shown in Fig. 7(b) and compared with that of the parent compound EuFe$_{2}$As$_{2}$ (Fig. 7(a)). Similar to the parent compound, the Eu $\mathit{\textrm{4}f}$ states in Eu(Fe$_{0.875}$Ir$_{0.125}$)$_{2}$As$_{2}$ are also quite localized, indicating that the Eu ions are in the stable 2+ valence state with a half filled $\mathit{\textrm{4}f}$ shell. Apart from the Eu $\mathit{\textrm{4}f}$ states, the remaining DOS changes significantly with Ir-doping. The enhancement of the band filling below the Fermi level (from 48.60 states/f.u. in EuFe$_{2}$As$_{2}$ to 49.24 states/f.u. in Eu(Fe$_{0.875}$Ir$_{0.125}$)$_{2}$As$_{2}$, as estimated by integrating the calculated total DOS below the Fermi level in Fig. 7(a) and (b), respectively) indicates that the substitution of Ir for Fe introduces electrons, similar to the effect of Co-doping. The total DOS at the Fermi level in Eu(Fe$_{0.875}$Ir$_{0.125}$)$_{2}$As$_{2}$ is 5.06 eV$^{-1}$ per unit cell, slightly reduced from 5.14 eV$^{-1}$ per unit cell in EuFe$_{2}$As$_{2}$. This decrease in the total DOS at the Fermi level is accompanied by the broadening of the $\mathit{d}$-band width caused by the Ir substitution. As shown in Fig. 7, for Eu(Fe$_{0.875}$Ir$_{0.125}$)$_{2}$As$_{2}$, the Fe 3$\mathit{d}$ band distributes throughout the range from -6.0 eV to 4.0 eV, more extended compared with that in the parent compound, which distributes from -5.5 eV to 3.8 eV. Meanwhile, the width of the As $\mathit{\textrm{4}p}$ band also increases, reflecting stronger $\mathit{d-p}$ hybridization due to the Ir doping. This might be attributed to the much more extended 5$\mathit{d}$ orbitals of Ir compared with the 3$\mathit{d}$ orbitals of Fe. The stronger hybridization and expanded bandwith is not favorable for Fermi surface nesting, and thus suppresses the structural distortion and the Fe-SDW transition leading to the emergence of superconductivity. This scenario is similar to that found for superconducting Sr(Fe$_{1-x}$Ir$_{x}$)$_{2}$As$_{2}$ by electronic structure calculations, where the suppression of the Fe-SDW order was attributed to the combined effects of the reduction in the Stoner enhancement, the increase in the bandwidth due to the hybridization involving Ir, and the additional introduction of electrons caused by the Ir substitution. \cite{ZhangLJ_09} In fact, further calculation indicates that in Eu(Fe$_{0.875}$Ir$_{0.125}$)$_{2}$As$_{2}$,
the Fe-SDW is indeed completely suppressed and the magnetic ground state of the Fe$^{2+}$ moments might be even ferromagnetic (with a very small moment of \textasciitilde{} 0.06 $\mathit{\mu_{B}}$), probably resulting from the proximity effect of the ferromagnetic
order in the Eu layers. 

\begin{figure}
\centering{}\includegraphics{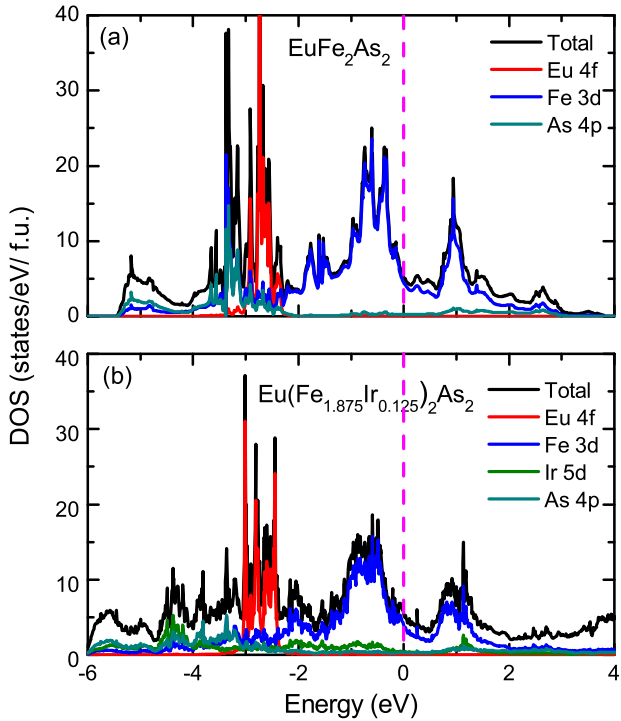}

\caption{The total and partial electronic density of states (DOS) per unit cell of EuFe$_{2}$As$_{2}$ (a) and Eu(Fe$_{0.875}$Ir$_{0.125}$)$_{2}$As$_{2}$ (b) in the quenched paramagnetic state in the Fe layer and the ferromagnetic interaction between the intralayer Eu spins in the Eu layer. The Fermi energy is set to zero (dashed line).}
\end{figure}

Energies of different possible magnetic structures of the Eu$^{2+}$ moments in Eu(Fe$_{0.875}$Ir$_{0.125}$)$_{2}$As$_{2}$ were also calculated and listed in Table II. Theoretically, the ferromagnetic alignment of the Eu$^{2+}$ spins along the $\mathit{c}$-direction
is indeed the most favored configuration, in good agreement with the observation presented in Section B. The magnetic order of the Eu$^{2+}$ spins does not play an important role in influencing the magnetism of the Fe sublattice and in contributing to the appearance of superconductivity. The two magnetic sublattices are almost decoupled. \cite{Xiao_09,Herrero-Martin_09,Jin_13}

\begin{table}
\caption{Energetic properties of the different Eu spin configurations for Eu(Fe$_{0.875}$Ir$_{0.125}$)$_{2}$As$_{2}$. The results are the total energy difference per Eu atom. }

\begin{ruledtabular} %
\begin{tabular}{cc}
 & Energy (meV)\tabularnewline
\hline 
A-AFM & 0.04641\tabularnewline
C-AFM  & 0.08825\tabularnewline
G-AFM & 0.08827\tabularnewline
FM along (001) & 0\tabularnewline
\end{tabular}\end{ruledtabular} 
\end{table}

\section{Discussion And Conclusion}

The coexistence of of superconductivity with a ferromagnetic ground state of the Eu$^{2+}$ moments seems to be a common feature of doped EuFe$_{2}$As$_{2}$, as it was universally observed in P-doped, \cite{Nandi_14} Co-doped, \cite{Jin_13}, Ru-doped, \cite{Jin_Ru}, and here in Ir-doped EuFe$_{2}$As$_{2}$. Although the role of dopants in suppressing the SPT and the Fe-SDW might be different to some extent, they tend to adjust and modify the indirect Ruderman-Kittel-Kasuya-Yosida (RKKY) interaction among the Eu$^{2+}$ spins in a very similar way, tuning the magnetic order of the Eu sublattice from A-type AFM in the non-superconducting parent compound to FM in the superconducting doped compound. It is intriguing that how the two antagonistic phenomena, superconductivity and ferromagnetism, can coexist in these compounds. As one possible solution of this puzzle, the existence of a spontaneous vortex state was suggested. \cite{Jiao_11} However, direct evidences for such a state are still lacking and additional measurements such as small angle neutron scattering (SANS) are needed.

In summary, the magnetic order of localized Eu$^{2+}$ spins in optimally-doped Eu(Fe$_{1-x}$Ir$_{x}$)$_{2}$As$_{2}$ ($\mathit{x}$ = 0.12) with superconducting transition temperature $\mathit{T_{SC}}$ = 22 K was investigated by single-crystal neutron diffraction. The Eu$^{2+}$ moments were found to be ferromagnetically aligned along the $\mathit{c}$-direction with an ordered moment of 7.0(1) $\mu_{B}$ well below the magnetic phase transition temperature $\mathit{T_{C}}$ = 17 K. The observed ordered moment is well consistent with the theoretical value of 7$\mu_{B}$ for an Eu$^{2+}$ ion. No evidence of the tetragonal-to-orthorhombic structural phase transition was found in this compound within the experimental uncertainty, in which the spin-density-wave (SDW) order of the Fe sublattice is supposed to be completely suppressed and the superconductivity is fully developed. The ferromagnetic ground state of the Eu$^{2+}$ spins in Eu(Fe$_{0.88}$Ir$_{0.12}$)$_{2}$As$_{2}$ is supported by first-principles magnetic structure calculations.
In addition, comparison of the electronic structure calculations between Eu(Fe$_{0.875}$Ir$_{0.125}$)$_{2}$As$_{2}$ and the parent compound EuFe$_{2}$As$_{2}$ indicates stronger hybridization and more expanded bandwith due to the Ir substitution, which together with the introduction of electrons might work against the Fe-SDW in favor of superconductivity. 

\begin{acknowledgments}
This work is based on experiments performed at the HEIDI instrument operated by Jülich Centre for Neutron Science (JCNS) at the Heinz Maier-Leibnitz Zentrum (MLZ), Garching, Germany. W. T. J. would like to acknowledge B. Schmitz and S. Mayr for the technical assistance, and K. Friese for helpful discussions. W. L. was supported by the Strategic Priority Research Program (B) of the Chinese Academy of Sciences (Grant No. XDB04010600), the National Natural Science Foundation of China (Grant No. 11404359), and the Shanghai Yang-Fan Program (Grant No. 14YF1407100). 
\end{acknowledgments}


\begin{thebibliography}{10}

\bibitem{Kamihara_08}
Y. Kamihara, T. Watanabe, M. Hirano, and H. Hosono, J. Am. Chem. Soc. \textbf{130}, 3296 (2008).

\bibitem{Cruz_08}
C. de la Cruz, Q. Huang, J. W. Lynn, J. Li, W. Ratcliff II, J. L. Zarestky, H. A. Mook, G. F. Chen, J. L. Luo, N. L. Wang, and P. Dai, Nature \textbf{453}, 899 (2008).	

\bibitem{Dai_12}
P. Dai, J. Hu, and E. Dagotto, Nat. Phys. \textbf{8}, 709 (2012).	

\bibitem{Marchand_78}
R. Marchand and W. Jeitschko, J. Solid State Chem. \textbf{24}, 351 (1978).	

\bibitem{Herrero-Martin_09}
J. Herrero-Mart\'in, V. Scagnoli, C. Mazzoli, Y. Su, R. Mittal, Y. Xiao, Th. Brueckel, N. Kumar, S. K. Dhar, A. Thamizhavel, and L. Paolasini, Phys. Rev. B \textbf{80}, 134411 (2009).	

\bibitem{Xiao_09}
Y. Xiao, Y. Su, M. Meven, R. Mittal, C. M. N. Kumar, T. Chatterji, S. Price, J. Person, N. Kumar, S. K. Dhar, A. Thamizhavel, and Th. Brueckel, Phys. Rev. B \textbf{80}, 174424 (2009).	

\bibitem{Jiang_09_NJP}
S. Jiang, Y. K. Luo, Z. Ren, Z. W. Zhu, C. Wang, X. F. Xu, Q. Tao, G. H. Cao, and Z. A. Xu, New. J. Phys. \textbf{11}, 025007 (2009).	

\bibitem{Jeevan_08}
H. S. Jeevan, Z. Hossain, D. Kasinathan, H. Rosner, C. Geibel, and P. Gegenwart, Phys. Rev. B \textbf{78}, 052502 (2008).	

\bibitem{Ren_09}
Z. Ren, Q. Tao, S. Jiang, C. Feng, C. Wang, J. Dai, G. Cao, and Z. Xu, Phys. Rev. Lett. \textbf{102}, 137002 (2009).	

\bibitem{Jiang_09}
S. Jiang, H. Xing, G. Xuan, Z. Ren, C. Wang, Z. A. Xu, and G. Cao, Phys. Rev. B \textbf{80}, 184514 (2009).	

\bibitem{Jiao_11}
W. H. Jiao, Q. Tao, J. K. Bao, Y. L. Sun, C. M. Feng, Z. A. Xu, I. Nowik, I. Feiner, and G. H. Cao, Europhys. Lett. \textbf{95}, 67007 (2011).	

\bibitem{Jiao_13}
W. H. Jiao, H. F. Zhai, J. K. Bao, Y. K. Luo, Q. Tao, C. M. Feng, Z. A. Xu, and G. H. Cao, New. J. Phys. \textbf{15}, 113002 (2013).	

\bibitem{Miclea_09}
C. F. Miclea, M. Nicklas, H. S. Jeevan, D. Kasinathan, Z. Hossain, H. Rosner, P. Gegenwart, C. Geibel, and F. Steglich, Phys. Rev. B \textbf{79}, 212509 (2009).	

\bibitem{Terashima_09}
T. Terashima,  M. Kimata,  H. Satsukawa, A. Harada, K. Hazama, S. Uji, H. S. Suzuki, T. Matsumoto, and K. Murata, J. Phys. Soc. Jpn. \textbf{78}, 083701 (2009).	

\bibitem{Jin_13}
W. T. Jin, S. Nandi, Y. Xiao, Y. Su, O. Zaharko, Z. Guguchia, Z. Bukowski, S. Price, W. H. Jiao, G. H. Cao, and Th. Brueckel, Phys. Rev. B \textbf{88}, 214516 (2013).	

\bibitem{Nandi_14}
S. Nandi, W. T. Jin, Y. Xiao, Y. Su, S. Price, D. K. Shukla, J. Strempfer, H. S. Jeevan, P. Gegenwart, and Th. Brueckel, Phys. Rev. B \textbf{89}, 014512 (2014).	

\bibitem{Nandi_14_neutron}
S. Nandi, W. T. Jin, Y. Xiao, Y. Su, S. Price, W. Schmidt, K. Schmalzl, T. Chatterji, H. S. Jeevan, P. Gegenwart, and Th. Brueckel, Phys. Rev. B \textbf{90}, 094407 (2014).	

\bibitem{Paramanik_13}
U. B. Paramanik, D. Das, R. Prasad, and Z. Hossain, J. Phys. Condens. Matter. \textbf{25}, 265701 (2013).

\bibitem{Xiao_10}
Y. Xiao, Y. Su, W. Schmidt, K. Schmalzl, C. M. N. Kumar, S. Price, T. Chatterji, R. Mittal, L. J. Chang, S. Nandi, et al., Phys. Rev. B (R) \textbf{81}, 220406 (2010).

\bibitem{Matin_07}
M. Meven, V. Hutanu, and G. Heger, Neutron News \textbf{18}, 19 (2007).

\bibitem{Coppens_65}
P. Coppens, L. Leiserowitz, and D. Rabinovich, Acta Crystallogr. \textbf{18}, 1035 (1965).

\bibitem{Rodriguez_93}
J. Rodr\'iguez-Carvajal, Physica B \textbf{192}, 55 (1993).	

\bibitem{Blochl_94}
P. E. Blöchl, Phys. Rev. B \textbf{50}, 17953 (1994).

\bibitem{Kresse_96}
G. Kresse and J. Furthmüller, Phys. Rev. B \textbf{54}, 11169 (1996).

\bibitem{Perdew_96}
J. P. Perdew, K. Burke, and M. Ernzerhof, Phys. Rev. Lett. \textbf{77}, 3865 (1996).

\bibitem{LiW_12}
W. Li, J. X. Zhu, Y. Chen, and C. S. Ting, Phys. Rev. B \textbf{86}, 155119 (2012).

\bibitem{Jeevan_11}
H. S. Jeevan, D. Kasinathan, H. Rosner, and P. Gegenwart, Phys. Rev. B \textbf{83}, 054511 (2011).	

\bibitem{Lester_09}
C. Lester, J.-H. Chu, J. G. Analytis, S. C. Capelli, A. S. Erickson, C. L. Condron, M. F. Toney, I. R. Fisher, and S. M. Hayden, Phys. Rev. B \textbf{79}, 144523 (2009).	

\bibitem{Footnote}
Note that the (0 0 6)\textsubscript{T} and (1 1 0)\textsubscript{T} reflections occur at comparable scattering vectors. The magnetic form factor of Eu\textsuperscript{2+} is 0.637 for (0 0 6)\textsubscript{T} and 0.785 for (1 1 0)\textsubscript{T}.

\bibitem{Fullprof_Manual}
Juan Rodr\'iguez-Carvajal, An introduction to the program FullProf 2000 (version 2001), Laboratoire L\'eon Brillouin (CEA-CNRS) (2011).

\bibitem{ZhangLJ_09}
L. J. Zhang and D. J. Singh, Phys. Rev. B \textbf{79}, 174530 (2009).

\bibitem{Jin_Ru}
W. T. Jin, et. al., unpublished.

\end{thebibliography}
\end{document}